\documentclass[sigconf,nonacm]{acmart}

\usepackage[acronym,toc]{glossaries}
\usepackage{algorithmic}
\usepackage{textcomp}
\usepackage{makecell}

\newcommand{\passagecolor}{black}
\newenvironment{addpassage}
{\color{\passagecolor}}
{}

\begin{document}


\title{Fairness in Multi-Agent Systems for Software Engineering: An SDLC-Oriented Rapid Review}

\author{Corey Yang-Smith}
\affiliation{%
  \institution{University of Calgary}
  \city{Calgary}
  \country{Canada}
}
\email{corey.yangsmith@ucalgary.ca}

\author{Ronnie de Souza Santos}
\affiliation{%
  \institution{University of Calgary}
  \city{Calgary}
  \country{Canada}
}
\email{ronnie.desouzasantos@ucalgary.ca}

\author{Ahmad Abdellatif}
\affiliation{%
  \institution{University of Calgary}
  \city{Calgary}
  \country{Canada}
}
\email{ahmad.abdellatif@ucalgary.ca}
\newacronym{llm}{LLM}{large language models}
\newacronym{mas}{MAS}{Multi-Agent Systems}
\newacronym{sdlc}{SDLC}{Software Development Life Cycle}
\newacronym{se}{SE}{Software Engineering}

\newcommand{\primaryPapers}{300}
\newcommand{\secondaryPapers}{50}
\newcommand{\initialPapers}{350}
\newcommand{\amountDedupe}{8}
\newcommand{\afterDedupe}{342}

\newcommand{\amountAbstractFilter}{313}
\newcommand{\afterAbstractFilter}{29}
\newcommand{\screeningFilter}{13}
\newcommand{\seedPapers}{16}

\newcommand{\snowballing}{2}
\newcommand{\finalPapers}{18}

\begin{abstract}
Transformer-based large language models (LLMs) and multi-agent systems (MAS) are increasingly embedded across the software development lifecycle (SDLC), yet their fairness implications for developer-facing tools remain underexplored despite their growing role in shaping what code is written, reviewed, and released. We present a rapid review of recent work on fairness in MAS, emphasizing LLM-enabled settings and relevance to software engineering. Starting from an initial set of 350 papers, we screened and filtered the corpus for relevance, retaining 18 studies for final analysis. Across these 18 studies, fairness is framed as a combination of trustworthy AI principles, bias reduction across groups, and interactional dynamics in collectives, while evaluation spans accuracy metrics on bias benchmarks, demographic disparity measures, and emergent MAS-specific notions such as conformity and bias amplification. Reported harms include representational, quality-of-service, security and privacy, and governance failures, which we relate to SDLC stages where evidence is most and least developed. We identify three persistent gaps: (1) fragmented, rarely MAS-specific evaluation practices that limit comparability, (2) limited generalization due to simplified environments and narrow attribute coverage, and (3) scarce, weakly evaluated mitigation and governance mechanisms aligned to real software workflows. These findings suggest MAS fairness research is not yet ready to support deployable, fairness-assured software systems, motivating MAS-aware benchmarks, consistent protocols, and lifecycle-spanning governance.
\end{abstract}



\maketitle

\section{Introduction}
Transformer-based \gls{llm} are increasingly used across the \gls{sdlc}, supporting tasks such as code generation \cite{Qian_2024_ChatDev}, testing \cite{Yu_2025_PATCHAGENT}, deployment \cite{ Mehta_2023_DevOps}, and maintenance \cite{Yang2025ASO}. The capabilities of these models have been further enhanced through research in reasoning and tool-use frameworks such as ReAct \cite{Yao_2023_ReAct} and chain-of-thought prompting \cite{Wei_2022_CoT}. Recent architectural trends increasingly favor \gls{mas}, in which multiple \gls{llm}-based agents collaborate on complex tasks \cite{He_2025_MultiAgent}, often integrating with external tools and platforms through emerging standards such as MCP \cite{MCP} and A2A \cite{A2A}. This trend expands their influence on the \gls{sdlc}, both in developer use and product integration, raising urgent questions about fairness, accountability, and responsible AI in software deployments.

Despite rapid advances in \gls{llm} research, systematic study of fairness in software engineering remains limited \cite{Xia_2025_LLMs}. Transparency challenges persist due to limited disclosure of training data, which can propagate bias into downstream tasks \cite{Pepe_2024_HFModels, Feng_2023_Pretraining_Bias}. AI assistants and low-code agentic tools are increasingly embedded in developer workflows and accessible to non-technical users \cite{Li_2025_Software30, githubCopilot, cursor2025, replit2025, n8n2025, zapier2025, openai2025agentkit}. As software engineering shifts towards AI-native development, often described as "Software 3.0" \cite{Hassan_2024_3.0, Hassan_2025_3.0}, agent-based architectures support collaboration, delegation, and autonomous execution alongside human developers. In this context, fairness extends beyond model outputs to include decision visibility, responsibility allocation, and human oversight, with implications for both developer workflows and downstream software products \cite{Ronanki_2025_Human_AI_Collaboration}.

This rapid review synthesizes current research on fairness in multi-agent software systems. After screening an initial set of 350 papers, retaining 18 studies for final analysis, we examine how fairness is defined and measured across existing \gls{llm}-based \gls{mas} studies, identify the types of harms and biases that arise within the \gls{sdlc}, and highlight where methodological or empirical gaps persist. This review contributes to responsible AI practice by offering guidance on how transparency, accountability, and equity can be strengthened as multi-agent architectures become increasingly explored and integrated into software engineering workflows and products. Our review focuses on three research questions:

\begin{itemize}
    \item \textbf{RQ1:} How is fairness defined and measured within existing studies of multi-agent systems?
    \item \textbf{RQ2:} What types of biases, harms, or inequitable outcomes have been identified across different stages of the \gls{sdlc} in these systems?
    \item \textbf{RQ3:} Where do current research gaps lie in promoting fairness, accountability, and transparency within multi-agent software ecosystems?

\end{itemize}

The remainder of this paper is organized as follows: Section~\ref{sec:background} reviews background and related work; Section~\ref{sec:methodology} describes the rapid review protocol; Section~\ref{sec:results} presents the results and synthesis across the three research questions; and Sections~\ref{sec:limitations} and ~\ref{sec:conclusion} discuss limitations and conclusions.

\section{Background and Related Work}
\label{sec:background}
\noindent\textbf{LLMs and MAS in Software Engineering.}
Since the introduction of the transformer \cite{Vaswani_2017_Attention}, \gls{llm}s have been applied across many \gls{se} tasks, including code generation, review, refactoring, and log analysis \cite{Hou_2024_LLM4SE_Lit_review}. Early work largely studied single-agent assistants spanning one or more \gls{sdlc} stages, evaluating how much a single model can support development and the resulting changes in reliability, maintainability, and control \cite{Jin_2025_LLM4SE_SAS}.

Recent research increasingly adopts multi-agent architectures that emulate software teams by assigning roles or coordinating agent pipelines \cite{Lin_2024_CodeGeneration_Multi-LLM}. Systems such as ChatDev \cite{Qian_2024_ChatDev} and MetaGPT \cite{Hong_2024_MetaGPT} implement waterfall-style processes while AgileCoder \cite{Nguyen_2024_AgileCoder} emphasizes iterative collaboration. Other work targets specific development activities such as IaC generation \cite{MACOG} and unit test synthesis and evaluation \cite{Xu_2025_CANDOR}. These \gls{mas}-based approaches align with AI-native "Software 3.0" development \cite{Hassan_2024_3.0, Hassan_2025_3.0}, but there remains a gap in understanding fairness, bias, and equitable treatment in \gls{mas}-supported \gls{se} \cite{Xia_2025_LLMs}.

\noindent\textbf{Fairness in LLM-Assisted Software Engineering.}
Fairness has emerged as an important but underexplored concern in applying AI to \gls{se}. Initial studies have shown that LLM generated code and technical suggestions can exhibit social biases toward certain demographic groups, for example by producing stereotyped identifiers, comments, or documentation \cite{Gallegos_2024_Bias, Kotek_2023_Gender_Bias}. Benchmark datasets such as MALIBU \cite{Mirza_2025_MALIBU_Bias_Benchmark} and BBQ \cite{Parrish_2022_BBQ_Dataset} reveal that persona-based \gls{llm} interactions can surface implicit gender, racial, and religious biases in model behavior. Other work focuses on real-world, human-centered coding scenarios, such as designing coding tasks and judgmental prompts that intentionally reveal latent biases in how LLMs complete or explain code \cite{Ling_2025_Social_Bias_in_Code}.

Within text-to-code settings, research by Liu \cite{Liu_2023_Social_Bias_in_Code} and Huang \cite{Huang_2025_Bias_in_LLM_Code} propose structured prompts, method signatures, and translation tasks to measure and mitigate bias in code synthesis and analysis. While these studies mainly address single model systems, they show that \gls{se} applications are subject to the same fairness challenges observed in natural language generation. They also highlight tradeoffs between bias reduction and downstream performance, described as an "alignment tax" by Xu \cite{Xu_2025_Social_Bias}, which are especially relevant when LLMs are embedded in developer tools. Beyond SE, a growing body of literature investigates fairness in MAS more generally, including:
\begin{itemize}
    \item \gls{mas} benchmarks evaluating implicit bias and differential treatment across personas or identity attributes \cite{Mirza_2025_MALIBU_Bias_Benchmark, Borah_2024_Implicit_Bias}
    \item Debate and deliberation frameworks that study group conformity \cite{Choi_2025_MAS_Group_Conformity}, political bias \cite{Bandaru_2025_Political_Bias_MAS}, and minority suppression in LLM-based social simulations \cite{Choi_2025_MAS_Group_Conformity}, and
    \item Conceptual and empirical work on responsibility allocation \cite{Ronanki_2025_Human_AI_Collaboration}, group blameworthiness \cite{Yazdanpanah_2021_Responsibility}, and collective misalignment in multi agent decision making \cite{Flint_2025_GroupSize}.
\end{itemize}
\begin{figure*}[t]
\centering
\setlength{\fboxsep}{4pt}
\setlength{\fboxrule}{0.4pt}

\fbox{%
  \parbox{\dimexpr\textwidth-2\fboxsep-2\fboxrule\relax}{%
    \ttfamily\footnotesize
    \setlength{\parindent}{0pt}%
    \setlength{\parskip}{0pt}%
    \linespread{0.92}\selectfont

    ("fairness" OR "bias" OR "equity" OR "justice" OR "discrimination" OR "harm" OR "accountability")
    {\normalfont\bfseries AND}
    ("multi-agent" OR "agentic system" OR "autonomous agent" OR "LLM-based agent" OR "multi-agent system" OR "MAS" OR "AI agent")
    {\normalfont\bfseries AND}
    ("software engineering" OR "software development lifecycle" OR "SDLC" OR "requirements engineering" OR "software design" OR "software architecture" OR "implementation" OR "testing" OR "software maintenance" OR "software evolution" OR "deployment" OR "operations" OR "MLOps" OR "DevOps")
  }%
}

\caption{Search strings to use across all databases in primary search.}
\vspace{-0.15cm}
\label{fig:search-strings}
\end{figure*}




While many of these studies are not grounded in \gls{se}, they propose definitions, metrics, and evaluation patterns that can be adapted to \gls{mas} embedded in software workflows and deployed within \gls{mas}-based software systems. For example, they introduce measures for bias amplification \cite{Nguyen_2025_Social_Cost}, group conformity \cite{Choi_2025_MAS_Group_Conformity}, and collective misalignment \cite{Flint_2025_GroupSize} that become relevant when multiple agents coordinate to propose designs, prioritize requirements, or vote on candidate patches.

\noindent\textbf{Responsible AI, Governance, and Trustworthiness Frameworks.}
Parallel to technical advances in \gls{llm}s and \gls{mas}, there is growing emphasis directed toward responsible AI frameworks and regulatory guidance. The EU AI Act \cite{EU_AI_Act_Website}, alongside related initiatives \cite{Ryan_2020_AI_Ethics}, formalize expectations around fairness, transparency, accountability, human oversight, and justice in AI systems. Recent studies \cite{decerqueira2025trustaiagentscase, Ronanki_2025_Human_AI_Collaboration, Raza_2025_TRiSM} operationalize these principles within software engineering contexts, outlining concrete requirements for AI-enabled workflows such as recruitment systems, DevOps automation, and human-in-the-loop collaboration. These works typically frame fairness as one dimension of trustworthy AI, alongside robustness, safety, privacy, and explainability. They discuss practices such as risk assessment, model documentation, continuous monitoring, and human-centric validation. However, they often focus on single systems or high-level governance rather than multi-agent architectures, and they rarely provide detailed fairness metrics or \gls{sdlc} applicable evaluation protocols.

Taken together, existing research on \gls{mas} in \gls{se}, fairness in LLM-based tools, and responsible AI governance reveals a fragmented landscape. Although multi-agent architectures are increasingly integrated into software workflows, fairness-focused analyses remain fragmented and are often disconnected from \gls{sdlc} practices. These disparities highlight the need for work that consolidates definitions, evaluation approaches, and fairness implications specific to \gls{mas} in \gls{se}.

\section{Methodology}
\label{sec:methodology}

This study followed a \textbf{rapid review} methodology as described by Cartaxo \cite{Cartaxo_2020_RapidReview}, prioritizing a timely synthesis by narrowing the search scope and using time-bounded screening of contemporary research, rather than the exhaustive coverage typical of traditional systematic literature reviews, as performed by other papers within the \gls{se}/\gls{llm} field \cite{Kawalerowicz, iwashima2025factorssupportgroundedresponses, Garcia2025}. This approach was chosen to capture recent work on fairness in multi-agent software systems, given the rapid pace of development in this area and the emerging need to address fairness concerns. The following subsections describe the procedures used throughout the study.

\begin{figure}[!b]
\centering
\vspace{-0.15cm}

\fbox{%
  \parbox{\columnwidth}{%
    \ttfamily\footnotesize
    \setlength{\parindent}{0pt}%
    \setlength{\parskip}{0pt}%
    \linespread{0.92}\selectfont
    "Multi-agent LLM" \textbf{AND} "Bias"
  }%
}

\caption{Search string used across Google Scholar during secondary review}
\label{fig:manual-search-strings}
\end{figure}

\subsection{Search Strategy}
\textbf{Sources and Time Span:} We searched the ACM Digital Library \cite{acmdl}, IEEE Xplore \cite{ieeexplore}, and Google Scholar \cite{googlescholar} for works published from 2017 to 2025. The search was conducted on November 15, 2025.

\textbf{Search Strategy:} We conducted a primary search across all databases using the string in Figure~\ref{fig:search-strings}, which was formed by combining fairness, multi-agent, and software engineering terms, retaining the top 100 results per database by relevance (300 total). We then ran a focused secondary search in Google Scholar using Figure~\ref{fig:manual-search-strings}, retaining 50 additional records.

\textbf{Snowballing:} One round of backward and forward snowballing was applied to seed papers identified through both searches to retrieve further relevant work.

\subsection{Inclusion, Exclusion and Screening Process}

After conducting the initial identification, papers were screened in multiple stages outlined below using the inclusion and exclusion criteria in Table~\ref{tab:screening-criteria}.
\begin{addpassage}
    We included both peer-reviewed conference and journal papers as well as preprints to capture recent developments in this area.    
\end{addpassage}
\begin{itemize}
    \item \textbf{Deduplication:} Remove duplicates and superseded versions.
    \item \textbf{Title and Abstract Screening:} Screen titles and abstracts to exclude irrelevant works; include peer-reviewed conference/journal papers and preprints.
    \item \textbf{Partial Text Screening:} Review Abstract, Introduction, Results, and Conclusion to assess relevance and record preliminary notes.
\end{itemize}

\begin{figure}[!t]
\centering
\captionsetup{skip=-6pt}
\includegraphics[width=\columnwidth, trim={0 8 0 8}, clip]{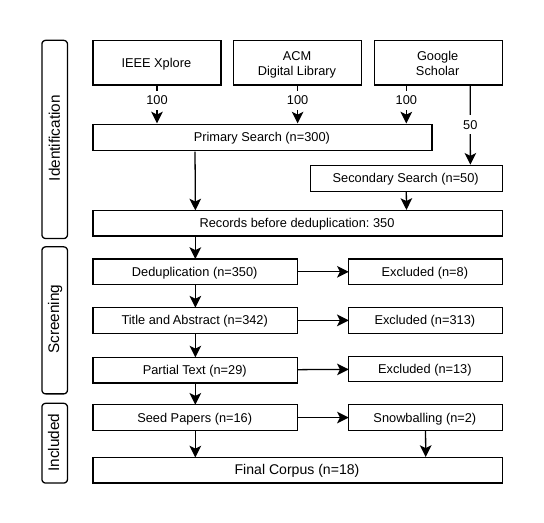}
\caption{Study selection flow for the rapid review.}

\vspace{-0.15cm}

\label{fig:screening_process}
\end{figure}

\textbf{Study Selection Results.} Figure~\ref{fig:screening_process} summarizes the number of records retained and excluded at each stage, resulting in \finalPapers{} included studies.

\subsection{Analysis Approach}
To address our research questions, we conducted a structural analysis of all studies that passed the screening process. The analysis was based on targeted data extraction and combined descriptive summarization with thematic synthesis. Our goal was to characterize how fairness is defined and assessed in multi-agent systems, identify documented harms across the \gls{sdlc}, and map remaining research gaps. For studies not explicitly situated in \gls{se}, we applied a predefined \gls{sdlc} mapping scheme that assigns reported harms and evaluation mechanisms to \gls{sdlc} stages based on their role in software workflows, rather than the original application domain.

\subsubsection{Information Extraction}
After study selection, we performed structured data extraction on each included paper to support the research questions. The first author skimmed the full text end-to-end, with particular attention to the Introduction, Methodology, Evaluation, Threats to Validity, and Future Work sections, revisiting specific parts as needed to clarify definitions, experimental setup, or findings related to fairness or multi-agent interactions. During the process, detailed notes were recorded in a spreadsheet\footnote{The coding spreadsheet used in this study is available at: \href{https://docs.google.com/spreadsheets/d/e/2PACX-1vQ-L9nAKYyAGir1kc9ZpJQy7k2tY2Gxc8Wj3JMwWC2aMA5VXzfM0xRAF9Flh_9PYpr9bKj2wIF-OY_7/pubhtml}{Google Sheets}} using a predefined extraction schema aligned with the research questions and organized into five categories:
\begin{itemize}
    \item \textbf{Fairness Definitions (RQ1):} How each study defined or conceptualized fairness in multi-agent contexts.
    \item \textbf{Fairness Evaluation Metrics (RQ1):} The metrics, criteria, or evaluation methods used to assess fairness or bias.
    \item \textbf{Harms and Inequities (RQ2):} The types of biases, harms, or inequitable outcomes reported or analyzed.
    \item \textbf{\gls{sdlc} Relevance (RQ2):} The stages of the \gls{sdlc} to which the harms, fairness concerns, or interventions applied.
    \item \textbf{Gaps and Future Work (RQ3):} Reported limitations, open problems, or directions for future research.
\end{itemize}

\begin{figure}[t]
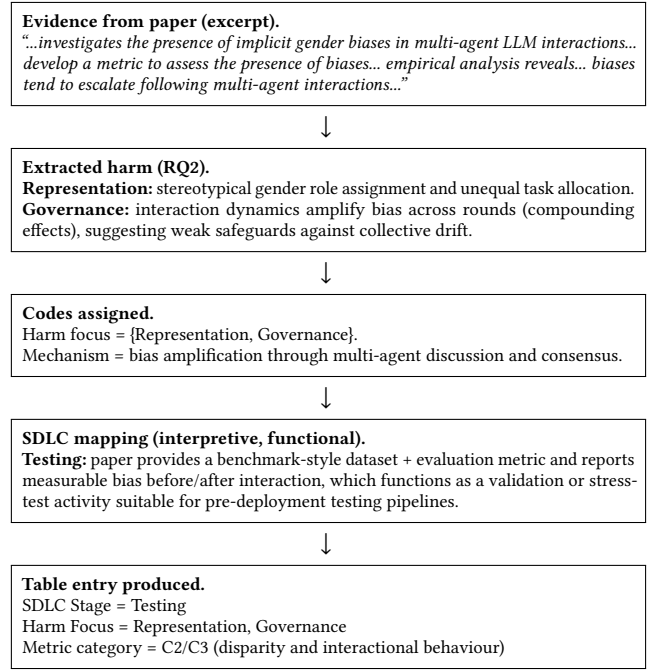

\centering
\footnotesize
\setlength{\fboxsep}{4pt}

\fbox{%
\begin{minipage}{0.96\columnwidth}
\textbf{Evidence from paper (excerpt).}\\
\emph{``...investigates the presence of implicit gender biases in multi-agent LLM interactions... develop a metric to assess the presence of biases... empirical analysis reveals... biases tend to escalate following multi-agent interactions...''}
\end{minipage}}

\vspace{1mm}
{\normalsize $\downarrow$}
\vspace{1mm}

\fbox{%
\begin{minipage}{0.96\columnwidth}
\textbf{Extracted harm (RQ2).}\\
\textbf{Representation:} stereotypical gender role assignment and unequal task allocation.\\
\textbf{Governance:} interaction dynamics amplify bias across rounds (compounding effects), suggesting weak safeguards against collective drift.
\end{minipage}}

\vspace{1mm}
{\normalsize $\downarrow$}
\vspace{1mm}

\fbox{%
\begin{minipage}{0.96\columnwidth}
\textbf{Codes assigned.}\\
Harm focus = \{Representation, Governance\}.\\
Mechanism = bias amplification through multi-agent discussion and consensus.
\end{minipage}}

\vspace{1mm}
{\normalsize $\downarrow$}
\vspace{1mm}

\fbox{%
\begin{minipage}{0.96\columnwidth}
\textbf{SDLC mapping (interpretive, functional).}\\
\textbf{Testing:} paper provides a benchmark-style dataset + evaluation metric and reports measurable bias before/after interaction, which functions as a validation or stress-test activity suitable for pre-deployment testing pipelines.
\end{minipage}}

\vspace{1mm}
{\normalsize $\downarrow$}
\vspace{1mm}

\fbox{%
\begin{minipage}{0.96\columnwidth}
\textbf{Table entry produced.}\\
SDLC Stage = Testing\\
Harm Focus = Representation, Governance\\
Metric category = C2/C3 (disparity and interactional behaviour)
\end{minipage}}

\caption{Coding example: how extracted evidence about implicit gender bias amplification is coded as representation and governance harm and mapped to the Testing stage via a functional SDLC interpretation \cite{Borah_2024_Implicit_Bias}.}
\vspace{-0.32cm}
\label{fig:coding}
\end{figure}

Considering the inclusion of non-\gls{se} studies, we treated a study as transferable to the \gls{se} domain when its agent interaction mechanism was domain-agnostic (e.g., debate, role assignment, consensus), its fairness failure mode was interaction-driven (e.g., amplification, conformity, minority suppression), and its evaluation could be instantiated on \gls{se} artifacts or workflows (e.g., requirements prioritization, design review, code review, test generation, or incident response). Figure~\ref{fig:coding} illustrates the analytical traceability from extracted study evidence to SDLC-aligned harms and themes using a representative example. Following extraction, we conducted iterative thematic synthesis by grouping recurring definitions, metrics, and reported harms across studies and refining themes until stable categories emerged that aligned with the research questions. These themes structure the synthesis presented in the Results and Discussion section. \begin{addpassage}
    Excluding preprints did not materially change the review's main findings, as the same high-level themes and research gaps remained visible in the peer-reviewed subset.
\end{addpassage}\begin{addpassage} Due to the rapid review scope, full independent double screening and coding were not feasible. A second author reviewed the extracted data and coding decisions for all included studies, and disagreements were resolved by consensus. This improved consistency, but does not replace formal inter-rater reliability assessment.\end{addpassage}

\begin{table}[b!]
\centering

\vspace{-0.32cm}

\caption{Screening criteria for study selection.}
\label{tab:screening-criteria}
\small
\renewcommand{\arraystretch}{1.20}
\setlength{\tabcolsep}{4pt}

\begin{tabular}{p{0.88\columnwidth}}
\toprule
\textbf{Include} \\
\midrule
Empirical evaluation or explicit conceptual/method contribution on multi-agent systems. \\
Fairness focus (bias, equity, discrimination, accountability) in multi-agent settings relevant to SE workflows or inter-agent interaction. \\
Proposes/evaluates fairness methods, metrics, frameworks, or benchmarks with transferable technical value to AI/SE. \\
English; 2017--2025. \\
\midrule
\textbf{Exclude} \\
\midrule
Non-SE applications where fairness analysis is not transferable to AI/SE practice. \\
Single-agent \gls{llm} studies without multi-agent interaction. \\
Non-technical ethics discussion without MAS details, evaluation, or SE implications. \\
Duplicates/superseded versions; full text unavailable. \\
\bottomrule
\end{tabular}
\end{table}

\subsubsection{Analysis Plan}
We applied a mixed-method analysis that combined structural coding with qualitative synthesis.

\textbf{RQ1: Fairness Conceptualization and Measurement.} We coded how fairness was defined or framed in each study and catalogued all reported evaluation metrics. These codes were then grouped to identify recurring conceptualizations, theoretical perspectives, and measurement practices. Labels were assigned based on the study's primary fairness objective stated in the framing and evaluation design. Metric categories (C1 to C4) were assigned based on the measurement used in the main evaluation; studies using multiple measurement types were assigned multiple categories.

\textbf{RQ2: Biases, Harms, and Inequitable Outcomes.} We analyzed reported harms and biases using thematic synthesis, grouping similar inequities and failure modes across studies. In parallel, we mapped each identified harm to relevant stages of the \gls{sdlc} to understand where fairness concerns arise within \gls{se} workflows. For studies not explicitly situated in \gls{se} contexts, this mapping was performed interpretively based on the functional role of the reported mechanism, evaluations, or interactions (e.g., testing, validation, or deployment), rather than the original application domain.

We mapped each harm or mechanism to \gls{sdlc} stages using decision rules: (i) governance, policy, or compliance constraints were coded as Requirements, (ii) mechanisms introduced as architectural or interaction controls were coded as Design; (iii) harms surfaced via benchmarks, stress tests, or disparity analyses were coded as Testing; (iv) and runtime amplification, drift, or operational failures were coded as Maintenance.

\textbf{RQ3: Research Gaps and Future Directions.} We analyzed limitations and future work statements across studies and clustered them into higher-level themes. These themes were then cross-referenced with findings from RQ1 and RQ2 to identify underexplored \gls{sdlc} stages, missing evaluation approaches, and inconsistencies in how fairness is assessed and operationalized.

This analysis produced both a conceptual characterization of fairness in \gls{mas} and a structural map of where literature is concentrated and where gaps remain.

\section{Results and Discussion}
\label{sec:results}
In this section, we synthesize findings from the included studies with respect to our three research questions. We highlight common patterns, points of divergence, research gaps, and implications for multi-agent software systems across the \gls{sdlc}. Table~\ref{tab:results-table} summarizes the included studies; the grouping rationale and synthesis dimensions used in the table are introduced in RQ1.


\begin{table*}[ht]
\small
\centering
\caption{Overview of sources grouped by fairness approach. Metric categories: C1 Benchmark Performance, C2 Group Disparity, C3 Interactional or MAS Behavioural, C4 Conceptual or Normative.}
\begin{tabular}{p{3.0cm}p{0.9cm}p{3.2cm}p{6.5cm}c}
\hline
Fairness Definition & Metrics & SDLC Stage & Harm Focus & Peer Reviewed \\
\hline
\multicolumn{5}{l}{\textbf{Reduce Bias ($n=9$)}} \\
Gosmar and Dahl \cite{Gosmar_2025_Sentinel_Agents} & C4 & Maintenance & Security, Quality of Service, Governance, Trust & -- \\
L\"unstedt and Schlippe \cite{Lunstedt_2025_Mitigating_Bias_in_MAS_Scenarios} & C1 & Design, Testing & Representation & \checkmark \\
Bandaru et al. \cite{Bandaru_2025_Political_Bias_MAS} & C2, C3 & Testing & Representation, Quality of Service, Governance & -- \\
Nguyen et al. \cite{Nguyen_2025_Social_Cost} & C1, C3 & N/A & Representation, Security & -- \\
Coppolillo et al. \cite{Coppolillo_2025_Conversational_Bias} & C2, C3 & Testing & Representation, Security, Quality of Service, Governance & -- \\
Xu et al. \cite{Xu_2025_Social_Bias} & C1 & N/A & Representation, Quality of Service & \checkmark \\
Solomon et al. \cite{Solomon_2025_LumiMAS} & C1 & Design, Testing & Security, Quality of Service & -- \\
Borah and Mihalcea \cite{Borah_2024_Implicit_Bias} & C2 & Testing & Representation, Governance & \checkmark \\
Mirza et al. \cite{Mirza_2025_MALIBU_Bias_Benchmark} & C1, C2 & Testing & Representation, Governance & \checkmark \\
\hline
\multicolumn{5}{l}{\textbf{\makecell[l]{Ethical and Trustworthy AI ($n=5$)}}}\\
Sharanarthi \cite{Sharanarthi_2025_MAS_Energy} & C4 & Development & Equity, Sustainability & \checkmark \\
Cerqueira et al. \cite{decerqueira2025trustaiagentscase} & C4 & Requirements, Development & Representation, Quality of Service, Governance, Trust & -- \\
Ronanki \cite{Ronanki_2025_Human_AI_Collaboration} & C4 & Requirements, Design & Governance, Trust & \checkmark \\
Oriol et al. \cite{Oriol_2025_MAD_RE} & C1 & Requirements & Quality of Service & \checkmark \\
Raza et al. \cite{Raza_2025_TRiSM} & C4 & Full SDLC & Security, Quality of Service, Governance & \checkmark \\
\hline
\multicolumn{5}{l}{\textbf{Inter-Agent Dynamics ($n=4$)}} \\
Choi et al. \cite{Choi_2025_Anonymization} & C3 & Testing & Representation, Governance, Trust & -- \\
Flint et al. \cite{Flint_2025_GroupSize} & C3 & Testing & Quality of Service & -- \\
Choi et al. \cite{Choi_2025_MAS_Group_Conformity} & C3 & Design, Testing & Quality of Service & \checkmark \\
Yazdanpanah et al. \cite{Yazdanpanah_2021_Responsibility} & C4 & Requirements & Governance & \checkmark \\
\hline
\end{tabular}

\vspace{-0.32cm}
\label{tab:results-table}
\end{table*}

\subsection{RQ1: Fairness Definitions and Measurement}

\textbf{Fairness Definitions.}
Across the multi-agent \gls{llm} literature, fairness tends to appear in three overlapping strands that reflect how authors define fairness goals and what they treat as evidence of progress. The first strand focuses on \textbf{reducing social bias} and group fairness in persona-based multi-agent simulations, where fairness means avoiding implicit or explicit stereotypes across protected groups such as gender \cite{Borah_2024_Implicit_Bias}, race, religion, disability, and politics \cite{Bandaru_2025_Political_Bias_MAS}. 
This line of work often relies on dedicated evaluation resources, such as the bias benchmark BBQ \cite{Parrish_2022_BBQ_Dataset}, including some proposed benchmarks such as MALIBU \cite{Mirza_2025_MALIBU_Bias_Benchmark}. 
Bias mitigation approaches such as MOMA \cite{Xu_2025_Social_Bias} aim to reduce bias.
\begin{addpassage}
Across the 18 included studies, gender was the most commonly examined attribute ($n=7$), followed by race ($n=6$), age ($n=4$), politics ($n=3$), disability ($n=3$), and intersectional bias ($n=1$).\end{addpassage}

A second strand ties fairness explicitly to normative and regulatory notions of justice, transparency, and non-discrimination, often referencing principles from the EU AI Act \cite{EU_AI_Act_Website} and ISO/IEC 42001:2023 \cite{ISO42001-2023} and considering fairness as a property of responsible human-agent collaboration and \textbf{ethical governance} rather than a purely statistical target \cite{decerqueira2025trustaiagentscase, Raza_2025_TRiSM, Ronanki_2025_Human_AI_Collaboration}. 

A third strand frames fairness through interactional and procedural dimensions within \textbf{multi-agent dynamics}. This view focuses on how responsibility and influence are distributed among agents \cite{Yazdanpanah_2021_Responsibility}, and how inter-agent dependency can contribute to unequal outcomes or propagate bias \cite{Flint_2025_GroupSize}. Flint et al. conceptualize fairness as the mitigation of political bias amplification during conversational debate, where agent viewpoints can converge and reinforce prejudice \cite{Flint_2025_GroupSize}. Choi et al. expands on this by treating fairness as the prevention of group conformity effects that accumulate over multiple interaction rounds \cite{Choi_2025_MAS_Group_Conformity}, with another study proposing agent anonymization as a prevention mechanism \cite{Choi_2025_Anonymization}.

\textbf{Fairness Measurement.}
Measurement choices align with the four metric categories summarized in Table~\ref{tab:results-table}.

\textbf{C1 (Benchmark Performance)} captures conventional predictive quality, where studies report accuracy, F1, or related scores on bias benchmarks such as BBQ \cite{Parrish_2022_BBQ_Dataset}. In this setting, improved benchmark performance is treated as a proxy for reduced bias or more equitable behavior \cite{Oriol_2025_MAD_RE, Lunstedt_2025_Mitigating_Bias_in_MAS_Scenarios}.

\textbf{C2 (Group Disparity)} measures fairness through between-group comparisons, examining whether scores, labels, or response distributions differ across protected attributes. These analyses may incorporate statistical testing to assess whether the observed gaps are significant \cite{Choi_2025_MAS_Group_Conformity} and, in some cases, rely on LLM-as-a-judge protocols \cite{Zheng_2023_LLM-as-a-Judge} to score outputs before computing disparities \cite{Coppolillo_2025_Conversational_Bias}.

\textbf{C3 (Interactional or MAS Behavioural)} reflects metrics tailored to multi-agent settings, where fairness concerns emerge from coordination dynamics rather than single-agent outputs. Representative approaches include measuring conformity or herding in debate and deliberation \cite{Choi_2025_MAS_Group_Conformity}, estimating collective misalignment in coordination problems \cite{Flint_2025_GroupSize}, and tracking propagated social cost or amplification effects over repeated interaction rounds \cite{Nguyen_2025_Social_Cost}.

Finally, \textbf{C4 (Conceptual or Normative)} is common in \gls{se}-oriented agent frameworks, where fairness is discussed as part of broader Trustworthy AI objectives such as explainability \cite{Sanwal_2025_Layered_CoT, Solomon_2025_LumiMAS}, security \cite{Gosmar_2025_Sentinel_Agents}, accountability, and sustainability \cite{Sharanarthi_2025_MAS_Energy}. While these works provide important design guidance, many of these studies do not define explicit fairness metrics, leaving a methodological gap and a lack of alignment between conceptual proposals and evaluation practices. 

\begin{addpassage}
Taken together, the three fairness strands identified not only define fairness differently but also measure it in largely non-overlapping ways: bias-reduction studies rely on benchmark accuracy and group disparity (C1/C2) but rarely assess interactional effects; inter-agent dynamics work introduces MAS-specific metrics such as conformity and amplification (C3) but seldom connects these to downstream demographic disparities; and governance-oriented studies discuss fairness normatively (C4) without specifying measurable thresholds. Because each strand evaluates a different facet of fairness using self-contained methods, results cannot be compared across strands, and no single study captures the full range of fairness concerns that arise when multiple agents coordinate within a software workflow.
\end{addpassage}
\subsection{RQ2: Biases and Harms Across the SDLC}
Across the reviewed work, relatively few studies directly examine LLM-based multi agent systems as developer-facing tools or as integrated components in software products. Instead, most evaluate \gls{mas} in general purpose settings such as social simulations, political discourse, or safety monitoring. Therefore, to answer RQ2 we first synthesize the harms identified in general \gls{mas} use, then map where these harms appear (or are likely to appear) across the \gls{sdlc}.

At a system level, the literature consistently reports several types of harm: representation harms (stereotypes, underrepresented viewpoints, and marginalization of protected groups), quality of service harms (unequal accuracy, hallucinations, misdiagnosis or unsafe recommendations), security and privacy harms (prompt injection, data exfiltration, and cascading attacks), and governance and trust challenges (opaque decision pathways, unclear responsibility, weak oversight). Several works also highlight bias amplification in multi agent interactions, where group conformity and persona reinforcement intensify existing bias \cite{Choi_2025_MAS_Group_Conformity}, as well as over-correction risks where mitigation strategies such as anonymization reduce accuracy or introduce new inequities \cite{Choi_2025_Anonymization}.
\begin{addpassage}
Among the reviewed \gls{mas} interaction patterns, two are particularly consequential: (i) iterative consensus through debate produces group conformity that progressively suppresses minority viewpoints, with severity scaling with group size and round count \cite{Choi_2025_MAS_Group_Conformity, Flint_2025_GroupSize}, while (ii) role-based task decomposition propagates stereotypical associations embedded in persona descriptions through the division of labor, causing downstream agents to inherit and amplify upstream biases \cite{Nguyen_2025_Social_Cost}.
\end{addpassage}

When mapped onto the \gls{sdlc}, these harms appear most predominantly at requirements, design, testing, and maintenance. Requirements engineering can encode or obscure ethical and legal constraints when \gls{mas} are used in domains such as recruitment or media integrity, raising representational harms if scenarios or prompts are biased \cite{decerqueira2025trustaiagentscase}. Architectural work treats fairness, security, and observability as cross-cutting design concerns, proposing role based access control, modular security and monitoring layers, and anonymization pipelines, each of which can shape downstream inequities \cite{Choi_2025_Anonymization}. Testing oriented benchmarks for conversational bias provide stress tests that could be integrated into pre-deployment pipelines, though they currently remain largely separate from mainstream  \gls{se} testing practice \cite{Xu_2025_Social_Bias}. Finally, security and observability frameworks position \gls{mas} as part of the operational fabric, where prompt injections, data leakage, hallucinations, and cascading failures surface in production logs or incident response \cite{Gosmar_2025_Sentinel_Agents}. Taken together, these studies outline a wide range of harms across the \gls{sdlc}, but the evidence remains uneven. Much of the work examines requirements, bias benchmarking, and security scenarios, while everyday developer activities such as design iteration, code review, and debugging of \gls{mas}-enabled tools receive limited attention.


\subsection{RQ3: Research Gaps and Implications for Fair MAS}
Although interest in multi-agent systems is accelerating, the evidence base remains hard to trust in practice because key claims are rarely tested under comparable agent-relevant conditions. Across the reviewed studies, shortcomings cluster around (i) evaluation that does not isolate agentic effects, (ii) coverage that is too configuration bound to generalize, and (iii) mitigations that are proposed more often than they are implemented.

\textbf{Evaluation Infrastructure.} A consistent limitation is that ``fairness" (and related accountability or transparency goals) is usually evaluated in ad-hoc settings that conflate agent interaction with task choice and evaluation design. Studies frequently vary core factors that directly shape group behaviour, such as the coordination protocol (debate, voting, delegation), agent count and role structure, memory and tool access, and prompting strategies. Many evaluations also rely on single tasks or general-purpose bias datasets that were not designed to probe interactional harms like conformity, polarization, dominance by a single agent, or bias amplification over multi-turn conversations. This makes cross-paper comparison difficult and weakens any architectural-level takeaway, as the differences in outcomes may reflect evaluation design rather than properties of multi-agent governance.

\textbf{Generalization.} A second gap is that results are often demonstrated in one narrow configuration (one foundation model, one interaction protocol, or one programming language), leaving it unclear which observed effects are intrinsic to multi-agent settings versus artifacts of the chosen setup. Coverage is also narrow in terms of harm dimensions: bias frequently centers on a small set of attributes (commonly gender or political preference), with limited attention to intersectional, linguistic, or domain-specific harms that are plausibly triggered by agent specialization and division of labor. Only one study in our corpus explicitly reports uneven performance across bias categories \cite{Lunstedt_2025_Mitigating_Bias_in_MAS_Scenarios}, and the literature rarely investigates \textit{why} these disparities emerge, for example whether they are driven by aggregation rules, role assignment, or task decomposition strategies. 

\textbf{Mitigation and Integration.} Finally, the corpus is richer in problem framing and risk identification than in deployable mitigations. Many papers surface interactional failure modes (bias amplification, group conformity, misaligned collective decisions), but few implement mitigations with ablations that separate ``agent effects" from ``prompting effects", compare alternatives under shared protocols, or evaluate practical constraints such as computing cost, scaling with agent count, and human oversight requirements. Without mitigations that are evaluated under \gls{mas}-relevant stressors and embedded into realistic \gls{se} pipelines, current progress remains closer to diagnosis than to operational governance for fair, accountable, and transparent multi-agent software ecosystems.


\section{Study Limitations}
\label{sec:limitations}

This rapid review is subject to several limitations that should be considered when interpreting the findings.

\textbf{Rapid Review Constraints:} As a rapid review, this study prioritizes timeliness over exhaustiveness. Searches were limited to selected databases, and pragmatic screening decisions were made under time and resource constraints, meaning some relevant studies may have been missed. We also included preprints to capture emerging work; however, their inclusion did not materially alter the main themes or conclusions. Since \gls{llm}-based \gls{mas} are evolving rapidly, this review should also be understood as a snapshot of the field at the time of study, and some conclusions may become outdated as new architectures, evaluation methods, and fairness interventions emerge. Therefore, the corpus is best interpreted as a well-motivated sample rather than a complete census. In addition, because the review focuses on \gls{llm}-based \gls{mas} in \gls{se}, the findings may not generalize to non-\gls{llm} \gls{mas} or to domains with different agent capabilities, interaction patterns, or fairness concerns.


\begin{addpassage}
    \textbf{Primary Author-Led Review Process:} Screening, data extraction, and initial coding were conducted primarily by the first author. Although a second author reviewed the extracted data and coding decisions for all included studies, it does not substitute for formal inter-rater-reliability assessment. Accordingly, some subjectivity may remain in the selection and classification of studies. Future works should incorporate more rigorous multi-coder procedures.
\end{addpassage}

\textbf{Temporal and Scope Limitations:} This review provides a snapshot of the field at a specific point in time. Because LLM-based \gls{mas} are evolving rapidly, new architectures, evaluation methods, and fairness interventions may emerge after the review window, and some conclusions may become outdated. Additionally, this study focuses on LLM-based \gls{mas} in \gls{se}; findings may not generalize to non-\gls{llm} \gls{mas} or to domains with substantially different agent capabilities, interaction patterns, or fairness concerns.

\begin{addpassage}
\textbf{Transferability from Non-SE Contexts:} A key limitation of this review is that many included studies are not grounded in \gls{se} contexts, instead focusing on general-purpose \gls{mas} settings such as social simulations or political discourse. Although we applied a functional role mapping, this approach primarily supports interaction-level generalization (e.g., bias amplification, conformity, coordination failures). In contrast, artifact-level outcomes that are central to \gls{se}, such as code quality, security vulnerabilities, and maintainability, may not be directly captured by these studies. As a result, our findings should be interpreted as identifying transferable fairness risks in multi-agent interaction patterns rather than providing empirical evidence of their impact on software artifacts or developer workflows. Future work should validate these interaction-driven risks in real-world \gls{se} settings.
\end{addpassage}

\section{Conclusion}
\label{sec:conclusion}
As foundation models and \gls{llm}-based multi-agent systems become embedded across the \gls{sdlc}, fairness must be evaluated as both an outcome property and an interactional system property. This rapid review synthesizes 18 studies and shows that \gls{mas} fairness work clusters into three strands (bias reduction, trustworthy AI governance, and inter-agent dynamics), but measurement remains fragmented across benchmark performance, group disparity, \gls{mas}-behavior metrics, and conceptual discussions that often lack operationalization. Reported harms span representation, quality of service, security and privacy, and governance and trust, with evidence concentrated in requirements, design, testing, and maintenance settings and comparatively limited attention to developer workflows such as code review and debugging. Overall, the literature is not yet positioned to support deployable fairness-assured \gls{mas} in \gls{se}, motivating \gls{mas}-aware benchmarks, consistent evaluation protocols, broader attribute and setting coverage, and mitigation and governance mechanisms validated in realistic pipelines.

\bibliographystyle{ACM-Reference-Format}
\bibliography{references}

@misc{ieeexplore,
  title        = {IEEE Xplore Digital Library},
  howpublished = {\url{https://ieeexplore.ieee.org}},
  note         = {Accessed: 2025-10-11}
}

@misc{acmdl,
  title        = {ACM Digital Library},
  howpublished = {\url{https://dl.acm.org/}},
  note         = {Accessed: 2025-10-11}
}

@misc{googlescholar,
  title        = {Google Scholar},
  howpublished = {\url{https://scholar.google.ca/}},
  note         = {Accessed: 2025-10-11}
}

@inproceedings{Yao_2023_ReAct,
  title = {{ReAct}: Synergizing Reasoning and Acting in Language Models},
  author = {Yao, Shunyu and Zhao, Jeffrey and Yu, Dian and Du, Nan and Shafran, Izhak and Narasimhan, Karthik and Cao, Yuan},
  booktitle = {International Conference on Learning Representations (ICLR) },
  year = {2023},
  html = {https://arxiv.org/abs/2210.03629},
}

@inproceedings{Wei_2022_CoT,
author = {Wei, Jason and Wang, Xuezhi and Schuurmans, Dale and Bosma, Maarten and Ichter, Brian and Xia, Fei and Chi, Ed H. and Le, Quoc V. and Zhou, Denny},
title = {Chain-of-thought prompting elicits reasoning in large language models},
year = {2022},
isbn = {9781713871088},
publisher = {Curran Associates Inc.},
address = {Red Hook, NY, USA},
booktitle = {Proceedings of the 36th International Conference on Neural Information Processing Systems},
articleno = {1800},
numpages = {14},
location = {New Orleans, LA, USA},
series = {NIPS '22}
}

@inproceedings{Vaswani_2017_Attention,
author = {Vaswani, Ashish and Shazeer, Noam and Parmar, Niki and Uszkoreit, Jakob and Jones, Llion and Gomez, Aidan N. and Kaiser, \L{}ukasz and Polosukhin, Illia},
title = {Attention is all you need},
year = {2017},
isbn = {9781510860964},
publisher = {Curran Associates Inc.},
address = {Red Hook, NY, USA},
booktitle = {Proceedings of the 31st International Conference on Neural Information Processing Systems},
pages = {6000–6010},
numpages = {11},
location = {Long Beach, California, USA},
series = {NIPS'17}
}

@misc{MCP,
  author       = {{Anthropic}},
  title        = {Introducing the Model Context Protocol},
  howpublished = {\url{https://www.anthropic.com/news/model-context-protocol}},
  year         = {2024},
  month        = nov,
  note         = {Accessed: \today}
}

@misc{A2A,
  author       = {Rao Surapaneni and Miku Jha and Michael Vakoc and Todd Segal},
  title        = {Announcing the Agent2Agent Protocol (A2A): A new era of Agent Interoperability},
  howpublished = {\url{https://developers.googleblog.com/en/a2a-a-new-era-of-agent-interoperability/}},
  year         = {2025},
  month        = apr,
  day          = {9},
  note         = {Accessed: \today}
}

@misc{Qian_2024_ChatDev,
      title={ChatDev: Communicative Agents for Software Development}, 
      author={Chen Qian and Wei Liu and Hongzhang Liu and Nuo Chen and Yufan Dang and Jiahao Li and Cheng Yang and Weize Chen and Yusheng Su and Xin Cong and Juyuan Xu and Dahai Li and Zhiyuan Liu and Maosong Sun},
      year={2024},
      eprint={2307.07924},
      archivePrefix={arXiv},
      primaryClass={cs.SE},
      url={https://arxiv.org/abs/2307.07924}, 
}

@misc{Hong_2024_MetaGPT,
      title={MetaGPT: Meta Programming for A Multi-Agent Collaborative Framework}, 
      author={Sirui Hong and Mingchen Zhuge and Jiaqi Chen and Xiawu Zheng and Yuheng Cheng and Ceyao Zhang and Jinlin Wang and Zili Wang and Steven Ka Shing Yau and Zijuan Lin and Liyang Zhou and Chenyu Ran and Lingfeng Xiao and Chenglin Wu and Jürgen Schmidhuber},
      year={2024},
      eprint={2308.00352},
      archivePrefix={arXiv},
      primaryClass={cs.AI},
      url={https://arxiv.org/abs/2308.00352}, 
}

@misc{Nguyen_2024_AgileCoder,
      title={AgileCoder: Dynamic Collaborative Agents for Software Development based on Agile Methodology}, 
      author={Minh Huynh Nguyen and Thang Phan Chau and Phong X. Nguyen and Nghi D. Q. Bui},
      year={2024},
      eprint={2406.11912},
      archivePrefix={arXiv},
      primaryClass={cs.SE},
      url={https://arxiv.org/abs/2406.11912}, 
}

@misc{MACOG,
      title={Multi-Agent Code-Orchestrated Generation for Reliable Infrastructure-as-Code}, 
      author={Rana Nameer Hussain Khan and Dawood Wasif and Jin-Hee Cho and Ali Butt},
      year={2025},
      eprint={2510.03902},
      archivePrefix={arXiv},
      primaryClass={cs.SE},
      url={https://arxiv.org/abs/2510.03902}, 
}

@misc{Xu_2025_CANDOR,
      title={Hallucination to Consensus: Multi-Agent LLMs for End-to-End Test Generation}, 
      author={Qinghua Xu and Guancheng Wang and Lionel Briand and Kui Liu},
      year={2025},
      eprint={2506.02943},
      archivePrefix={arXiv},
      primaryClass={cs.SE},
      url={https://arxiv.org/abs/2506.02943}, 
}

@inbook{Yu_2025_PATCHAGENT,
author = {Yu, Zheng and Guo, Ziyi and Wu, Yuhang and Yu, Jiahao and Xu, Meng and Mu, Dongliang and Chen, Yan and Xing, Xinyu},
title = {PATCHAGENT: a practical program repair agent mimicking human expertise},
year = {2025},
isbn = {978-1-939133-52-6},
publisher = {USENIX Association},
address = {USA},
articleno = {226},
numpages = {20}
}

@misc{Mehta_2023_DevOps,
      title={Automated DevOps Pipeline Generation for Code Repositories using Large Language Models}, 
      author={Deep Mehta and Kartik Rawool and Subodh Gujar and Bowen Xu},
      year={2023},
      eprint={2312.13225},
      archivePrefix={arXiv},
      primaryClass={cs.SE},
      url={https://arxiv.org/abs/2312.13225}, 
}

@misc{githubCopilot,
  author       = {{GitHub, Inc.}},
  title        = {GitHub Copilot: Your AI Pair Programmer},
  howpublished = {\url{https://github.com/features/copilot}},
  year         = {2025},
  note         = {Accessed: \today}
}

@misc{replit2025,
  author       = {{Replit}},
  title        = {Replit — Build apps and sites with AI},
  howpublished = {\url{https://replit.com/}},
  year         = {2025},
  note         = {Accessed: \today}
}

@misc{cursor2025,
  author       = {{Cursor}},
  title        = {Cursor — The AI Code Editor},
  howpublished = {\url{https://cursor.com/}},
  year         = {2025},
  note         = {Accessed: \today}
}

@misc{zapier2025,
  author       = {{Zapier Inc.}},
  title        = {Zapier: Automate AI Workflows, Agents, and Apps},
  howpublished = {\url{https://zapier.com/}},
  year         = {2025},
  note         = {Accessed: \today}
}

@misc{n8n2025,
  author       = {{n8n}},
  title        = {n8n — AI Workflow Automation Platform \& Tools},
  howpublished = {\url{https://n8n.io/}},
  year         = {2025},
  note         = {Accessed: \today}
}

@misc{openai2025agentkit,
  author       = {{OpenAI}},
  title        = {Introducing AgentKit},
  howpublished = {\url{https://openai.com/index/introducing-agentkit/}},
  year         = {2025},
  month        = oct,
  day          = {6},
  note         = {Accessed: \today}
}

@misc{Hassan_2024_3.0,
      title={Towards AI-Native Software Engineering (SE 3.0): A Vision and a Challenge Roadmap}, 
      author={Ahmed E. Hassan and Gustavo A. Oliva and Dayi Lin and Boyuan Chen and Zhen Ming Jiang},
      year={2024},
      eprint={2410.06107},
      archivePrefix={arXiv},
      primaryClass={cs.SE},
      url={https://arxiv.org/abs/2410.06107}, 
}

@misc{Hassan_2025_3.0,
      title={Agentic Software Engineering: Foundational Pillars and a Research Roadmap}, 
      author={Ahmed E. Hassan and Hao Li and Dayi Lin and Bram Adams and Tse-Hsun Chen and Yutaro Kashiwa and Dong Qiu},
      year={2025},
      eprint={2509.06216},
      archivePrefix={arXiv},
      primaryClass={cs.SE},
      url={https://arxiv.org/abs/2509.06216}, 
}

@inproceedings{Pepe_2024_HFModels,
author = {Pepe, Federica and Nardone, Vittoria and Mastropaolo, Antonio and Bavota, Gabriele and Canfora, Gerardo and Di Penta, Massimiliano},
title = {How do Hugging Face Models Document Datasets, Bias, and Licenses? An Empirical Study},
year = {2024},
isbn = {9798400705861},
publisher = {Association for Computing Machinery},
address = {New York, NY, USA},
url = {https://doi.org/10.1145/3643916.3644412},
doi = {10.1145/3643916.3644412},
booktitle = {Proceedings of the 32nd IEEE/ACM International Conference on Program Comprehension},
pages = {370–381},
numpages = {12},
location = {Lisbon, Portugal},
series = {ICPC '24}
}

@misc{Xia_2025_LLMs,
      title={Analyzing 16,193 LLM Papers for Fun and Profits}, 
      author={Zhiqiu Xia and Lang Zhu and Bingzhe Li and Feng Chen and Qiannan Li and Chunhua Liao and Feiyi Wang and Hang Liu},
      year={2025},
      eprint={2504.08619},
      archivePrefix={arXiv},
      primaryClass={cs.DL},
      url={https://arxiv.org/abs/2504.08619}, 
}

@misc{Cartaxo_2020_RapidReview,
      title={Rapid Reviews in Software Engineering}, 
      author={Bruno Cartaxo and Gustavo Pinto and Sergio Soares},
      year={2020},
      eprint={2003.10006},
      archivePrefix={arXiv},
      primaryClass={cs.SE},
      url={https://arxiv.org/abs/2003.10006}, 
}

@inproceedings{Feng_2023_Pretraining_Bias,
    title = "From Pretraining Data to Language Models to Downstream Tasks: Tracking the Trails of Political Biases Leading to Unfair {NLP} Models",
    author = "Feng, Shangbin  and
      Park, Chan Young  and
      Liu, Yuhan  and
      Tsvetkov, Yulia",
    editor = "Rogers, Anna  and
      Boyd-Graber, Jordan  and
      Okazaki, Naoaki",
    booktitle = "Proceedings of the 61st Annual Meeting of the Association for Computational Linguistics (Volume 1: Long Papers)",
    month = jul,
    year = "2023",
    address = "Toronto, Canada",
    publisher = "Association for Computational Linguistics",
    url = "https://aclanthology.org/2023.acl-long.656/",
    doi = "10.18653/v1/2023.acl-long.656",
    pages = "11737--11762",
}

@misc{Li_2025_Software30,
      title={The Rise of AI Teammates in Software Engineering (SE) 3.0: How Autonomous Coding Agents Are Reshaping Software Engineering}, 
      author={Hao Li and Haoxiang Zhang and Ahmed E. Hassan},
      year={2025},
      eprint={2507.15003},
      archivePrefix={arXiv},
      primaryClass={cs.SE},
      url={https://arxiv.org/abs/2507.15003}, 
}

@article{He_2025_MultiAgent,
author = {He, Junda and Treude, Christoph and Lo, David},
title = {LLM-Based Multi-Agent Systems for Software Engineering: Literature Review, Vision, and the Road Ahead},
year = {2025},
issue_date = {June 2025},
publisher = {Association for Computing Machinery},
address = {New York, NY, USA},
volume = {34},
number = {5},
issn = {1049-331X},
url = {https://doi.org/10.1145/3712003},
doi = {10.1145/3712003},
journal = {ACM Trans. Softw. Eng. Methodol.},
month = may,
articleno = {124},
numpages = {30}
}

@inproceedings{Choi_2025_MAS_Group_Conformity,
    title = "An Empirical Study of Group Conformity in Multi-Agent Systems",
    author = "Choi, Min  and
      Kim, Keonwoo  and
      Chae, Sungwon  and
      Baek, Sangyeop",
    editor = "Che, Wanxiang  and
      Nabende, Joyce  and
      Shutova, Ekaterina  and
      Pilehvar, Mohammad Taher",
    booktitle = "Findings of the Association for Computational Linguistics: ACL 2025",
    month = jul,
    year = "2025",
    address = "Vienna, Austria",
    publisher = "Association for Computational Linguistics",
    url = "https://aclanthology.org/2025.findings-acl.265/",
    doi = "10.18653/v1/2025.findings-acl.265",
    pages = "5123--5139",
    ISBN = "979-8-89176-256-5",
}

@misc{Mirza_2025_MALIBU_Bias_Benchmark,
      title={MALIBU Benchmark: Multi-Agent LLM Implicit Bias Uncovered}, 
      author={Imran Mirza and Cole Huang and Ishwara Vasista and Rohan Patil and Asli Akalin and Sean O'Brien and Kevin Zhu},
      year={2025},
      eprint={2507.01019},
      archivePrefix={arXiv},
      primaryClass={cs.CL},
      url={https://arxiv.org/abs/2507.01019}, 
}

@misc{decerqueira2025trustaiagentscase,
      title={Can We Trust AI Agents? A Case Study of an LLM-Based Multi-Agent System for Ethical AI}, 
      author={José Antonio Siqueira de Cerqueira and Mamia Agbese and Rebekah Rousi and Nannan Xi and Juho Hamari and Pekka Abrahamsson},
      year={2025},
      eprint={2411.08881},
      archivePrefix={arXiv},
      primaryClass={cs.CY},
      url={https://arxiv.org/abs/2411.08881}, 
}

@misc{Bandaru_2025_Political_Bias_MAS,
      title={Revealing Political Bias in LLMs through Structured Multi-Agent Debate}, 
      author={Aishwarya Bandaru and Fabian Bindley and Trevor Bluth and Nandini Chavda and Baixu Chen and Ethan Law},
      year={2025},
      eprint={2506.11825},
      archivePrefix={arXiv},
      primaryClass={cs.AI},
      url={https://arxiv.org/abs/2506.11825}, 
}

@misc{Flint_2025_GroupSize,
      title={Group size effects and collective misalignment in LLM multi-agent systems}, 
      author={Ariel Flint and Luca Maria Aiello and Romualdo Pastor-Satorras and Andrea Baronchelli},
      year={2025},
      eprint={2510.22422},
      archivePrefix={arXiv},
      primaryClass={cs.MA},
      url={https://arxiv.org/abs/2510.22422}, 
}

@misc{Nguyen_2025_Social_Cost,
      title={The Social Cost of Intelligence: Emergence, Propagation, and Amplification of Stereotypical Bias in Multi-Agent Systems}, 
      author={Thi-Nhung Nguyen and Linhao Luo and Thuy-Trang Vu and Dinh Phung},
      year={2025},
      eprint={2510.10943},
      archivePrefix={arXiv},
      primaryClass={cs.MA},
      url={https://arxiv.org/abs/2510.10943}, 
}

@misc{Coppolillo_2025_Conversational_Bias,
      title={Unmasking Conversational Bias in AI Multiagent Systems}, 
      author={Erica Coppolillo and Giuseppe Manco and Luca Maria Aiello},
      year={2025},
      eprint={2501.14844},
      archivePrefix={arXiv},
      primaryClass={cs.CL},
      url={https://arxiv.org/abs/2501.14844}, 
}

@misc{Choi_2025_Anonymization,
      title={Measuring and Mitigating Identity Bias in Multi-Agent Debate via Anonymization}, 
      author={Hyeong Kyu Choi and Xiaojin Zhu and Sharon Li},
      year={2025},
      eprint={2510.07517},
      archivePrefix={arXiv},
      primaryClass={cs.AI},
      url={https://arxiv.org/abs/2510.07517}, 
}

@inbook{Ronanki_2025_Human_AI_Collaboration,
author = {Ronanki, Krishna},
title = {Facilitating Trustworthy Human-Agent Collaboration in LLM-based Multi-Agent System oriented Software Engineering},
year = {2025},
isbn = {9798400712760},
publisher = {Association for Computing Machinery},
address = {New York, NY, USA},
url = {https://doi.org/10.1145/3696630.3728717},
booktitle = {Proceedings of the 33rd ACM International Conference on the Foundations of Software Engineering},
pages = {1333–1337},
numpages = {5}
}

@INPROCEEDINGS {Oriol_2025_MAD_RE,
author = { Oriol, Marc and Motger, Quim and Marco, Jordi and Franch, Xavier },
booktitle = { 2025 IEEE 33rd International Requirements Engineering Conference (RE) },
title = {{ Multi-Agent Debate Strategies to Enhance Requirements Engineering with Large Language Models }},
year = {2025},
volume = {},
ISSN = {},
pages = {527-534},
doi = {10.1109/RE63999.2025.00063},
url = {https://doi.ieeecomputersociety.org/10.1109/RE63999.2025.00063},
publisher = {IEEE Computer Society},
address = {Los Alamitos, CA, USA},
month =sep}

@misc{Raza_2025_TRiSM,
      title={TRiSM for Agentic AI: A Review of Trust, Risk, and Security Management in LLM-based Agentic Multi-Agent Systems}, 
      author={Shaina Raza and Ranjan Sapkota and Manoj Karkee and Christos Emmanouilidis},
      year={2025},
      eprint={2506.04133},
      archivePrefix={arXiv},
      primaryClass={cs.AI},
      url={https://arxiv.org/abs/2506.04133}, 
}

@article{
Xu_2025_Social_Bias, title={Mitigating Social Bias in Large Language Models: A Multi-Objective Approach Within a Multi-Agent Framework}, volume={39}, url={https://ojs.aaai.org/index.php/AAAI/article/view/34748}, DOI={10.1609/aaai.v39i24.34748},number={24}, journal={Proceedings of the AAAI Conference on Artificial Intelligence}, author={Xu, Zhenjie and Chen, Wenqing and Tang, Yi and Li, Xuanying and Hu, Cheng and Chu, Zhixuan and Ren, Kui and Zheng, Zibin and Lu, Zhichao}, year={2025}, month={Apr.}, pages={25579-25587} }

@ARTICLE{Yazdanpanah_2021_Responsibility,
  author={Yazdanpanah, Vahid and Gerding, Enrico H. and Stein, Sebastian and Cirstea, Corina and Schraefel, M. C. and Norman, Timothy J. and Jennings, Nicholas R.},
  journal={IEEE Internet Computing}, 
  title={Different Forms of Responsibility in Multiagent Systems: Sociotechnical Characteristics and Requirements}, 
  year={2021},
  volume={25},
  number={6},
  pages={15-22},
  doi={10.1109/MIC.2021.3107334}}

@misc{Solomon_2025_LumiMAS,
      title={LumiMAS: A Comprehensive Framework for Real-Time Monitoring and Enhanced Observability in Multi-Agent Systems}, 
      author={Ron Solomon and Yarin Yerushalmi Levi and Lior Vaknin and Eran Aizikovich and Amit Baras and Etai Ohana and Amit Giloni and Shamik Bose and Chiara Picardi and Yuval Elovici and Asaf Shabtai},
      year={2025},
      eprint={2508.12412},
      archivePrefix={arXiv},
      primaryClass={cs.CR},
      url={https://arxiv.org/abs/2508.12412}, 
}

@misc{Borah_2024_Implicit_Bias,
      title={Towards Implicit Bias Detection and Mitigation in Multi-Agent LLM Interactions}, 
      author={Angana Borah and Rada Mihalcea},
      year={2024},
      eprint={2410.02584},
      archivePrefix={arXiv},
      primaryClass={cs.CL},
      url={https://arxiv.org/abs/2410.02584}, 
}

@inproceedings{Kotek_2023_Gender_Bias, series={CI ’23},
   title={Gender bias and stereotypes in Large Language Models},
   url={http://dx.doi.org/10.1145/3582269.3615599},
   DOI={10.1145/3582269.3615599},
   booktitle={Proceedings of The ACM Collective Intelligence Conference},
   publisher={ACM},
   author={Kotek, Hadas and Dockum, Rikker and Sun, David},
   year={2023},
   month=nov, pages={12–24},
   collection={CI ’23} }

@misc{Parrish_2022_BBQ_Dataset,
      title={BBQ: A Hand-Built Bias Benchmark for Question Answering}, 
      author={Alicia Parrish and Angelica Chen and Nikita Nangia and Vishakh Padmakumar and Jason Phang and Jana Thompson and Phu Mon Htut and Samuel R. Bowman},
      year={2022},
      eprint={2110.08193},
      archivePrefix={arXiv},
      primaryClass={cs.CL},
      url={https://arxiv.org/abs/2110.08193}, 
}

@misc{Ling_2025_Social_Bias_in_Code,
      title={Bias Unveiled: Investigating Social Bias in LLM-Generated Code}, 
      author={Lin Ling and Fazle Rabbi and Song Wang and Jinqiu Yang},
      year={2025},
      eprint={2411.10351},
      archivePrefix={arXiv},
      primaryClass={cs.SE},
      url={https://arxiv.org/abs/2411.10351}, 
}

@misc{Huang_2025_Bias_in_LLM_Code,
      title={Bias Testing and Mitigation in LLM-based Code Generation}, 
      author={Dong Huang and Jie M. Zhang and Qingwen Bu and Xiaofei Xie and Junjie Chen and Heming Cui},
      year={2025},
      eprint={2309.14345},
      archivePrefix={arXiv},
      primaryClass={cs.SE},
      url={https://arxiv.org/abs/2309.14345}, 
}

@misc{Liu_2023_Social_Bias_in_Code,
      title={Uncovering and Quantifying Social Biases in Code Generation}, 
      author={Yan Liu and Xiaokang Chen and Yan Gao and Zhe Su and Fengji Zhang and Daoguang Zan and Jian-Guang Lou and Pin-Yu Chen and Tsung-Yi Ho},
      year={2023},
      eprint={2305.15377},
      archivePrefix={arXiv},
      primaryClass={cs.CL},
      url={https://arxiv.org/abs/2305.15377}, 
}

@misc{Hou_2024_LLM4SE_Lit_review,
      title={Large Language Models for Software Engineering: A Systematic Literature Review}, 
      author={Xinyi Hou and Yanjie Zhao and Yue Liu and Zhou Yang and Kailong Wang and Li Li and Xiapu Luo and David Lo and John Grundy and Haoyu Wang},
      year={2024},
      eprint={2308.10620},
      archivePrefix={arXiv},
      primaryClass={cs.SE},
      url={https://arxiv.org/abs/2308.10620}, 
}

@misc{Jin_2025_LLM4SE_SAS,
      title={From LLMs to LLM-based Agents for Software Engineering: A Survey of Current, Challenges and Future}, 
      author={Haolin Jin and Linghan Huang and Haipeng Cai and Jun Yan and Bo Li and Huaming Chen},
      year={2025},
      eprint={2408.02479},
      archivePrefix={arXiv},
      primaryClass={cs.SE},
      url={https://arxiv.org/abs/2408.02479}, 
}

@misc{Lin_2024_CodeGeneration_Multi-LLM,
      title={SOEN-101: Code Generation by Emulating Software Process Models Using Large Language Model Agents}, 
      author={Feng Lin and Dong Jae Kim and Tse-Husn Chen},
      year={2024},
      eprint={2403.15852},
      archivePrefix={arXiv},
      primaryClass={cs.SE},
      url={https://arxiv.org/abs/2403.15852}, 
}

@misc{EU_AI_Act_Website,
  title        = {{EU Artificial Intelligence Act: Up-to-date developments and analyses}},
  howpublished = {\url{https://artificialintelligenceact.eu/}},
  note         = {Accessed: 2025-11-29}
}

@article{Ryan_2020_AI_Ethics,
    author = {Ryan, Mark and Stahl, Bernd Carsten},
    title = {Artificial intelligence ethics guidelines for developers and users: clarifying their content and normative implications},
    journal = {Journal of Information, Communication and Ethics in Society},
    volume = {19},
    number = {1},
    pages = {61-86},
    year = {2020},
    month = {06},
    issn = {1477-996X},
    doi = {10.1108/JICES-12-2019-0138},
    url = {https://doi.org/10.1108/JICES-12-2019-0138},
    eprint = {https://www.emerald.com/jices/article-pdf/19/1/61/1616450/jices-12-2019-0138.pdf},
}

@standard{ISO42001-2023,
  title        = {Information technology — Artificial intelligence — Management system},
  author       = {{International Organization for Standardization (ISO) and IEC}},
  year         = {2023},
  number       = {ISO/IEC 42001:2023},
  edition      = {1},
  month        = dec,
  organization = {ISO / IEC},
  note         = {Published December 2023; 51 pages}
}

@misc{Zheng_2023_LLM-as-a-Judge,
      title={Judging LLM-as-a-Judge with MT-Bench and Chatbot Arena}, 
      author={Lianmin Zheng and Wei-Lin Chiang and Ying Sheng and Siyuan Zhuang and Zhanghao Wu and Yonghao Zhuang and Zi Lin and Zhuohan Li and Dacheng Li and Eric P. Xing and Hao Zhang and Joseph E. Gonzalez and Ion Stoica},
      year={2023},
      eprint={2306.05685},
      archivePrefix={arXiv},
      primaryClass={cs.CL},
      url={https://arxiv.org/abs/2306.05685}, 
}

@article{Gallegos_2024_Bias,
    title = "Bias and Fairness in Large Language Models: A Survey",
    author = "Gallegos, Isabel O.  and
      Rossi, Ryan A.  and
      Barrow, Joe  and
      Tanjim, Md Mehrab  and
      Kim, Sungchul  and
      Dernoncourt, Franck  and
      Yu, Tong  and
      Zhang, Ruiyi  and
      Ahmed, Nesreen K.",
    journal = "Computational Linguistics",
    volume = "50",
    number = "3",
    month = sep,
    year = "2024",
    address = "Cambridge, MA",
    publisher = "MIT Press",
    url = "https://aclanthology.org/2024.cl-3.8/",
    doi = "10.1162/coli_a_00524",
    pages = "1097--1179",
}

@INPROCEEDINGS{Sharanarthi_2025_MAS_Energy,
  author={Sharanarthi, Tanush},
  booktitle={2025 5th International Symposium on Computer Technology and Information Science (ISCTIS)}, 
  title={Adaptive Multi-Agent AI Framework for Real-Time Energy Optimization and Context-Aware Code Review in Software Development}, 
  year={2025},
  volume={},
  number={},
  pages={353-358},
  doi={10.1109/ISCTIS65944.2025.11066037}}

@INPROCEEDINGS{Lunstedt_2025_Mitigating_Bias_in_MAS_Scenarios,
  author={Lünstedt, Jens and Schlippe, Tim},
  booktitle={2025 7th International Conference on Natural Language Processing (ICNLP)}, 
  title={Mitigating Bias in Large Language Models Leveraging Multi-Agent Scenarios}, 
  year={2025},
  volume={},
  number={},
  pages={14-18},
  doi={10.1109/ICNLP65360.2025.11108428}}

@misc{Sanwal_2025_Layered_CoT,
      title={Layered Chain-of-Thought Prompting for Multi-Agent LLM Systems: A Comprehensive Approach to Explainable Large Language Models}, 
      author={Manish Sanwal},
      year={2025},
      eprint={2501.18645},
      archivePrefix={arXiv},
      primaryClass={cs.CL},
      url={https://arxiv.org/abs/2501.18645}, 
}

@misc{Gosmar_2025_Sentinel_Agents,
      title={Sentinel Agents for Secure and Trustworthy Agentic AI in Multi-Agent Systems}, 
      author={Diego Gosmar and Deborah A. Dahl},
      year={2025},
      eprint={2509.14956},
      archivePrefix={arXiv},
      primaryClass={cs.AI},
      url={https://arxiv.org/abs/2509.14956}, 
}

@article{Yang2025ASO,
  title={A Survey of LLM-based Automated Program Repair: Taxonomies, Design Paradigms, and Applications},
  author={Boyang Yang and Zijian Cai and Feng Liu and Bach Le and Lingming Zhang and T{\'e}gawend{\'e} F. Bissyand{\'e} and Yang Liu and Haoye Tian},
  journal={ArXiv},
  year={2025},
  volume={abs/2506.23749},
  url={https://api.semanticscholar.org/CorpusID:280010745}
}

@misc{iwashima2025factorssupportgroundedresponses,
      title={Factors That Support Grounded Responses in LLM Conversations: A Rapid Review}, 
      author={Gabriele Cesar Iwashima and Claudia Susie Rodrigues and Claudio Dipolitto and Geraldo Xexéo},
      year={2025},
      eprint={2511.21762},
      archivePrefix={arXiv},
      primaryClass={cs.CL},
      url={https://arxiv.org/abs/2511.21762}, 
}

@InProceedings{Kawalerowicz,
author="Kawalerowicz, Marcin
and Pietranik, Marcin
and St{\k{e}}pniak, Krzysztof",
editor="Nguyen, Ngoc Thanh
and Dinh Duc Anh, Vu
and Kozierkiewicz, Adrianna
and Nguyen Van, Sinh
and N{\'u}{\~{n}}ez, Manuel
and Treur, Jan
and Vossen, Gottfried",
title="LLMs as Code Review Agents: A Rapid Review and Experimental Evaluation with Human Expert Judges",
booktitle="Computational Collective Intelligence",
year="2026",
publisher="Springer Nature Switzerland",
address="Cham",
pages="346--360",
isbn="978-3-032-09318-9"
}

@article{Garcia2025,
  author  = {Garcia, Manuel B.},
  title   = {Teaching and learning computer programming using ChatGPT: A rapid review of literature amid the rise of generative AI technologies},
  journal = {Education and Information Technologies},
  year    = {2025},
  volume  = {30},
  number  = {12},
  pages   = {16721--16745},
  issn    = {1573-7608},
  doi     = {10.1007/s10639-025-13452-5},
  url     = {https://doi.org/10.1007/s10639-025-13452-5},
}
\end{document}